\documentclass[runningheads]{llncs}
\usepackage[T1]{fontenc}
\usepackage{bm}
\usepackage{multirow}
\usepackage{graphicx}
\usepackage{hyperref}
\usepackage{amsmath}

\usepackage{amsthm}
\usepackage{algorithmic}  
\usepackage{amsmath}
\usepackage{booktabs}
\usepackage{array}
\usepackage{caption}
\newcolumntype{C}[1]{>{\centering\arraybackslash}p{#1}}
\usepackage{float} 
\usepackage{stfloats}
\usepackage{hyperref}
\usepackage{tabularx}
\usepackage{subcaption}
\usepackage{amsmath}
\usepackage{amssymb}

\begin{document}
\title{Label Inference Attacks against \\Federated Unlearning}
%
\institute{}
\author{Wei Wang\inst{1}\orcidID{0009-0009-0625-4603} \and
Xiangyun Tang\inst{1}\thanks {The corresponding authors: Xiangyun Tang (xiangyunt@muc.edu.cn), Yajie Wang
(wangyajie0312@foxmail.com).  This work was supported in part by the National Nature Science Foundation of China under Grant 62302539, 62402040 and the Key Laboratory of Ethnic Language Intelligent Analysis and Security Management of MOE (OPU-202502)  }\orcidID{0000-0002-5511-0720} \and \\
Yajie Wang\inst{2}$^{\star}$\orcidID{0000-0002-7023-4844} \and
Yijing Lin\inst{3}\orcidID{0000-0003-2702-7679} \and \\
Tao Zhang\inst{4}\orcidID{0000-0002-2639-4357} \and
Meng Shen\inst{2}\orcidID{0000-0001-5706-6383} \and \\
Dusit Niyato\inst{5}\orcidID{0000-0002-7442-7416} \and
Liehuang Zhu\inst{2}\orcidID{0000-0003-3277-3887}
}

\authorrunning{W. Wang et al.}
%
\institute{
the Key Laboratory of Ethnic Language Intelligent Analysis and Security Management of MOE, Minzu University of China, Beijing, China 
\email{\{wangwei,xiangyunt\}@muc.edu.cn}\\
\and School of Cyberspace Science and Technology, Beijing Institute of Technology, Beijing, China\\
\email{wangyajie0312@foxmail.com}\\
\and School of Information and Communication Engineering, Beijing University of Posts and Telecommunications, Beijing, China\\
\and School of Computer Science and Engineering, Beijing Jiaotong University, Beijing, China\\
\and School of Cyberspace Science and Technology, Nanyang Technological University, Singapore\\
}
\maketitle              

\vspace{-0.3cm}
\begin{abstract}
Federated Unlearning (FU) has emerged as a promising solution to respond to ``the right to be forgotten” of clients, by allowing clients to erase their data from global models without compromising model performance. 
Unfortunately, researchers find that the parameter variations of models induced by FU expose clients’ data information, enabling attackers to infer the label of unlearning data, while label inference attacks against FU remain unexplored.
In this paper, we introduce and analyze a new privacy threat against FU and propose a novel label inference attack, $\mathtt{ULIA}$, which can infer unlearning data labels across three FU levels.
To address the unique challenges of inferring labels via
the models variations, we design a gradient-label mapping mechanism in $\mathtt{ULIA}$ that establishes a relationship between gradient variations and unlearning labels, enabling inferring labels on accumulated model variations. We evaluate $\mathtt{ULIA}$ on both IID and non-IID settings.
Experimental results show that in the IID setting, $\mathtt{ULIA}$ achieves a $100\%$ Attack Success Rate (ASR) under both class-level and client-level unlearning. 
Even when only $1\%$ of a user's local data is forgotten, $\mathtt{ULIA}$ still attains an ASR ranging from $93\%$ to $62.3\%$. 

\end{abstract}
\section{Introduction}
Federated Learning (FL), as a decentralized machine learning paradigm, has gained widespread adoption in various domains such as finance \cite{long2020federated} and smart cities \cite{jiang2020federated} due to its inherent capability to protect user privacy.
FL allows multiple users to jointly train a global model by sharing local models rather than raw data with a central server, thereby mitigating privacy leakage \cite{mothukuri2021survey}.
In addition to the privacy protection, existing data security legislation, such as General Data Protection Regulation (GDPR) \cite{voigt2017eu} and California Consumer Privacy Act (CCPA) \cite{harding2019understanding}, emphasizes the ``Right to be Forgotten", affirming users' authority to demand the unlearning of their data from global models during FL.

To respond the unlearning demand, Federated Unlearning (FU) has emerged as a promising solution \cite{liu2024survey}.
FU methods, such as historical information-based unlearning \cite{zhao2023federated} and rapid retraining \cite{liu2022right}, involve the server collaborating with users to remove the influence of unlearning data from the global model, in accordance with users' unlearning requests, while ensuring that the model's performance remains consistent with its state prior to the unlearning operation. 
FU can be categorized into sample-level, class-level, and client-level \cite{sheng2024robust}, where the unlearning requests correspond to a set of samples, all samples associated with a set of classes, or the entire local dataset of a user, respectively.

Although existing FU methods can unlearn data from global models while preserving model performance, they are vulnerable to a significant privacy leakage threat. 
After FU operations, the server holds two versions of models before and after unlearning. 
The adjustments of the resulting parameter on the models are not entirely independent but are closely related to the characteristics of the unlearning data, such as the labels of the unlearning data \cite{shaik2024exploring}. 
Hence, \textit{the variations in the models before and after FU expose users' data information}, enabling the server as attackers to infer private information about the unlearning data.
In this paper, we focus on label inference attacks against FU, where the server, by analyzing the variations of models before and after unlearning, infers the labels of unlearning data. 

Label inference attacks allow the server to steal private labels of users that should have been unlearning or protected in FU, raising significant privacy concerns.
Furthermore, ensuring the privacy of labels is a fundamental guarantee, as they often represent sensitive or critical information for the participant \cite{qiu2022your}. 
For example, in a healthcare scenario, multiple medical institutions collaboratively train a global model using diabetic patient data. 
If a patient requests the removal of their medical data due to privacy concerns, the label inference attacks on FU enable the attacker to infer the diagnosis, leading to a severe breach of privacy.

However, the privacy risks of label inference attacks on FU remain underexplored.
It is challenging to infer the labels of unlearning data from the model differences induced by FU. 
The model differences reflect the accumulated impact of all the unlearning data and their associated labels. 
Consequently, when attackers are unaware of the number of unlearning data samples or labels, accurately inferring labels from the accumulated parameter differences is challenging.
Furthermore, when the quantity of unlearning data is small, the resulting changes in model parameters may be too subtle to adequately capture the features of the unlearning data, making it more difficult for attackers to infer the labels.

In this paper, we propose $\mathtt{ULIA}$, a novel label inference attack that infers the labels of unlearning data across three FU levels: sample-level, class-level, and client-level, by analyzing the variations in the model parameters of FU.
$\mathtt{ULIA}$ addresses the inherent challenge of accurately inferring labels from model differences induced by FU, where the accumulated impact of unlearning data on model parameters obscures individual label effects. 
To overcome this, we propose a \textit{gradient-label mapping mechanism}, which establishes a relationship between gradient variations and unlearning labels.
This allows the attacker to separate the specific parameter shifts attributable to individual unlearning label, thus enhancing the accuracy of label inference. 
Moreover, $\mathtt{ULIA}$ incorporates a dynamic filtering strategy that prioritizes label categories with the higher likelihood of matching model parameter changes, focusing on those labels that exhibit the most significant alterations, ensuring effective label inference, even in scenarios with sparse or subtle unlearning data. 

We evaluate the performance of $\mathtt{ULIA}$ under three advanced FU methods across the three FU levels on real-world datasets. In the IID setting, $\mathtt{ULIA}$ achieves an Attack Success Rate (ASR) ranging from $100\%$ to $62.3\%$. Even in the non-IID setting, $\mathtt{ULIA}$ is still able to achieve $96.4\% - 58.5\%$ ASR.

The main contributions of this paper are as follows:
\begin{itemize} 
\item To the best of our knowledge, we are the first to reveal the label leakage issue of FU. We propose $\mathtt{ULIA}$, a novel attack inferring unlearning data labels across three FU levels, by analyzing the variations in the model parameters induced by unlearning operations.

\item The attack works regardless of the quantity of unlearning data, as we design a gradient-label mapping mechanism that establishes a relationship between gradient variations and unlearning labels, enabling inferring unlearning labels on accumulated model variations.

\item We evaluate our attacks with real-world datasets under three advanced FU methods, both in IID and non-IID settings. The experimental results show that $\mathtt{ULIA}$ demonstrates outstanding attack performance and adaptability.
\end{itemize}

\section{Related Works}
\vspace{-0.3cm}
\subsubsection{Federated Unlearning.}
Existing FU methods can be broadly categorized into two main approaches:
(1) \textit{Historical information-based unlearning}, which is recorded and analyzed during training to assess the impact of specific data or clients on the global model, enables efficient unlearning. 
The typical techniques explored under this approach include: 
Gradient correction adjusts the gradient information in the model training process to eliminate the contribution of target data or target clients to the global model \cite{zhao2023federated}. 
The knowledge distillation strategy restores the performance of the global model through knowledge transfer and model tuning, thereby approximating unlearning \cite{wu2022federated}. 
Approximate unlearning of target data is achieved by pruning network layers \cite{wang2022federated} and refining the loss function \cite{jiang2024efficient}. 
(2) \textit{Rapid retraining}, which efficiently restores the unlearning global model by optimizing retraining algorithms or performing partial retraining. 
The typical techniques explored under this approach include:
First-order Taylor expansion methods are employed to effectively utilize gradient and curvature information, thereby identifying a more optimal descent direction \cite{liu2022right}. 
The optimal number of unlearning rounds can be accurately determined by assessing the contribution of target clients during each training iteration \cite{lin2024blockchain}. 
Clients are grouped into suitable clusters for aggregation, ensuring that retraining occurs solely within the cluster of the target client during the unlearning process \cite{su2023asynchronous}.

\vspace{-0.3cm}
\subsubsection{Label Inference Attacks.}
Label inference attacks aim to infer the labels of training or unlearned data, thereby compromising the privacy of participating clients. 
Previous studies have primarily implemented label inference attacks through the following techniques. 
(1) \textit{Based on gradient information}, where attackers exploit the label information reflected in the gradient signs and magnitudes to perform inference. \cite{zhao2020idlg} discovers the relationship between labels and gradient signs in the cross-entropy loss function, which becomes a fundamental basis for label inference attacks. 
\cite{fu2022label} demonstrates that attackers exploit inherent vulnerabilities in vertical federated learning (VFL) to infer sensitive labels owned by one party through the output features of the bottom model or the gradient information returned by the server. 
(2) \textit{Classification and clustering methods}, where attackers train gradient classifiers or utilize the clustering properties of embeddings and gradients to make inferences. 
\cite{li2021label} observes that in two-party scenarios, the gradient norm associated with target labels is often larger than that of non-target labels, providing a basis for classification-based methods.
\cite{liu2022clustering} proposes a K-means clustering-based attack method that classifies gradients or embeddings using cosine similarity. 
(3) \textit{Model reconstruction}, where attackers simulate the predictive model and labels of the active party to construct surrogate models and labels, minimizing prediction errors to infer the true labels. 
\cite{zhang2022data} proposes using label smoothing techniques to prevent the model from becoming overconfident in label predictions. 
\cite{arazzi2023blindsage} introduces an early stopping strategy to reduce the impact of gradient magnitude peaks on the attack accuracy.

\vspace{-0.3cm}
\subsubsection{Highlights of the proposed attack.}
Existing label inference attacks mainly focus on inferring the labels of training data. 
With the emergence of FU, the analysis of local and global model parameters' changes generated by unlearning requests to infer the labels of forgotten data has still remained unexplored. 
In this paper, we propose $\mathtt{ULIA}$, a novel label inference attack against FU, inferring unlearning data labels by analyzing the model variations induced by unlearning operations, which reveals and analyzes the new privacy leakage issue of FU.

\section{Preliminaries}
Before entering the sample-level FU process, clients participate in FL, training global models under the server’s orchestration. 
The FU process is launched when a client launches an unlearning request. The server and clients cooperate to eliminate the target data from the global model, based on a unlearning strategy.

\vspace{-0.3cm}
\subsubsection{Federated Learning.}
The primary objective of FL is to enable collaborative and efficient training on distributed data without directly sharing the raw data. Given $N$ clients $P_i \ (i = 1, \ldots, N)$, each holding local data $D_i$, the overall dataset is defined as $D = \bigcup_{i=1}^N D_i$
. At the beginning of the FL training process, the server initializes the global model $\theta_{\mathrm{global}}^{(0)}$ and distributes it to all clients. At the $t$-th round, client $i$ optimizes the global model $\theta_{\mathrm{global}}^{(t)}$ based on its local data $D_i$, and updates the local model parameters $\theta_{\mathrm{local},i}^{(t+1)}$:
\begin{equation}
\theta_{\mathrm{local},i}^{(t+1)}=\theta_{\mathrm{global}}^{\left(t\right)}-\eta \cdot \sum_{(x, y) \in D_i} \nabla \ell(f_{\theta}(x), y)
\end{equation}
where $\eta$ is the learning rate, and $\nabla \ell(f_{\theta}(x), y)$ is the gradient of the loss function $\ell$ for model $f_{\theta}$ with respect to input $x$ and label $y$.

The server aggregates the local model parameters $\theta_{\mathrm{local},i}^{(t+1)}$ uploaded by all participating after their individual training steps, in order to update the global model parameters $\theta_{\mathrm{global}}^{(t+1)}$:
\begin{equation}
\theta_{\mathrm{global}}^{\left(t+1\right)} = \sum_{i=1}^{N} w_i \theta_{\mathrm{local},i}^{\left(t+1\right)}
\label{eq:global_update}
\end{equation}
where \( w_i = \frac{|D_i|}{|D|} \) represents the weight of client \( i \) based on the size of its local dataset \( D_i \) relative to the total dataset \( D \).
\vspace{-0.3cm}
\subsubsection{Federated Unlearning.}
When a target client \( K \) sends an unlearning request in the \( t \)-th round, FU performs the data removal request, resulting in the post-unlearning global model $\theta_{\text{global}}'$. FU can be categorized into the following three main types based on different forgetting requests and objectives. 
\begin{itemize}  
    \item \textit{Sample-level unlearning} seeks to remove specific sensitive samples from a client's local dataset and eliminate their influence on the global model to protect privacy \cite{zhang2023fedrecovery}. The local dataset $D_i$ is updated by removing $S_f$, resulting in a revised dataset $D_i' = D_i \setminus S_f$. The client initializes its post-unlearning local model $\theta_{\mathrm{local}, i}'$ with the global model parameters $\theta_{\text{global}}^{(t)}$ from the $t$-th round and retrains it on the updated dataset $D_i'$. 
    \item
    \textit{Class-level unlearning} focuses on completely erasing certain class-specific information from the global model, renders the model incapable of classifying those classes \cite{che2023fast,wang2024efficient}. When the influence of a specific class $C_f$ needs to be removed, each client updates its local dataset by removing the data associated with $C_f$, resulting in modified datasets $D_i' = D_i \setminus \{(x, y) \mid y \in C_f\}$. 
    \item \textit{Client-level unlearning} aims to fully remove the contributions of specific clients and to ensure that their data has no residual impact on the global model or subsequent training processes \cite{yuan2023federated}. The global model removes the influence of the target clients $K$ and updates using only the remaining clients. The remaining dataset is represented as $D' = \bigcup_{\substack{i = 1 \\ i \neq K}}^{N} D_i$.
\end{itemize}

\section{Attack Model and Problem Formalization}
\subsection{Attack Model}
This paper focuses on label inference attacks against FU. When a target client submits an unlearning request to remove samples from the global model or to exit collaborative training, the server and client participate in FU to ensure that the model no longer reflects the influence of the forgotten data. We assume that the server is semi-honest. 
By exploiting access to the model parameters, the semi-honest server as the attacker aims to infer the labels of the data that the target client has requested to forget. 
\vspace{-0.3cm}
\subsubsection{Attacker’s Goal.} 
The attacker's goal is to infer the labels of the unlearning data by analyzing the parameter differences of models before and after unlearning. Our attack specifically targets a single client’s unlearning request but considers different types of forgotten data.
\vspace{-0.3cm}
\subsubsection{Attacker's Knowledge.}
The semi-honest server, acting as the attacker, possesses legitimate data within FU. It can access two versions of the local model parameters and global model parameters, before and after unlearning. 

\subsection{Problem Formalization}
Throughout this paper, we study the problem of label inference attacks against FU. We denote the forgotten dataset as \({D}_{\text{forgotten}} = \{(x_i, y_i) \mid i = 1, \dots, M\}\), where \(x_i\) represents the input features, and \(y_i \in \{0, 1\}^C\) denotes the corresponding one-hot encoded class labels for \(C\) classes. Let the global model \(\mathcal{F}(\cdot; \theta)\) be parameterized by \(\theta\), which maps the input space \(\mathcal{X}\) to a \(C\)-dimensional output space \(\mathbb{R}^C\). As a potential attacker, the server analyzes parameter differences between the two versions of both the local and the global models to infer the labels of the forgotten data, formally expressed as the following objective function:
\begin{equation}
\hat{y}_i = \arg\max_{y_i} \mathcal{L}\left(\mathcal{F}(x_i; \theta_{\text{global}}), y_i \right)
\end{equation}
where \(\hat{y}_i\) is the inferred label for the forgotten data point \(x_i\), and \(\mathcal{L}\) is the loss measuring the discrepancy between the model output and the predicted label.

\begin{figure}
\includegraphics[width=\textwidth]{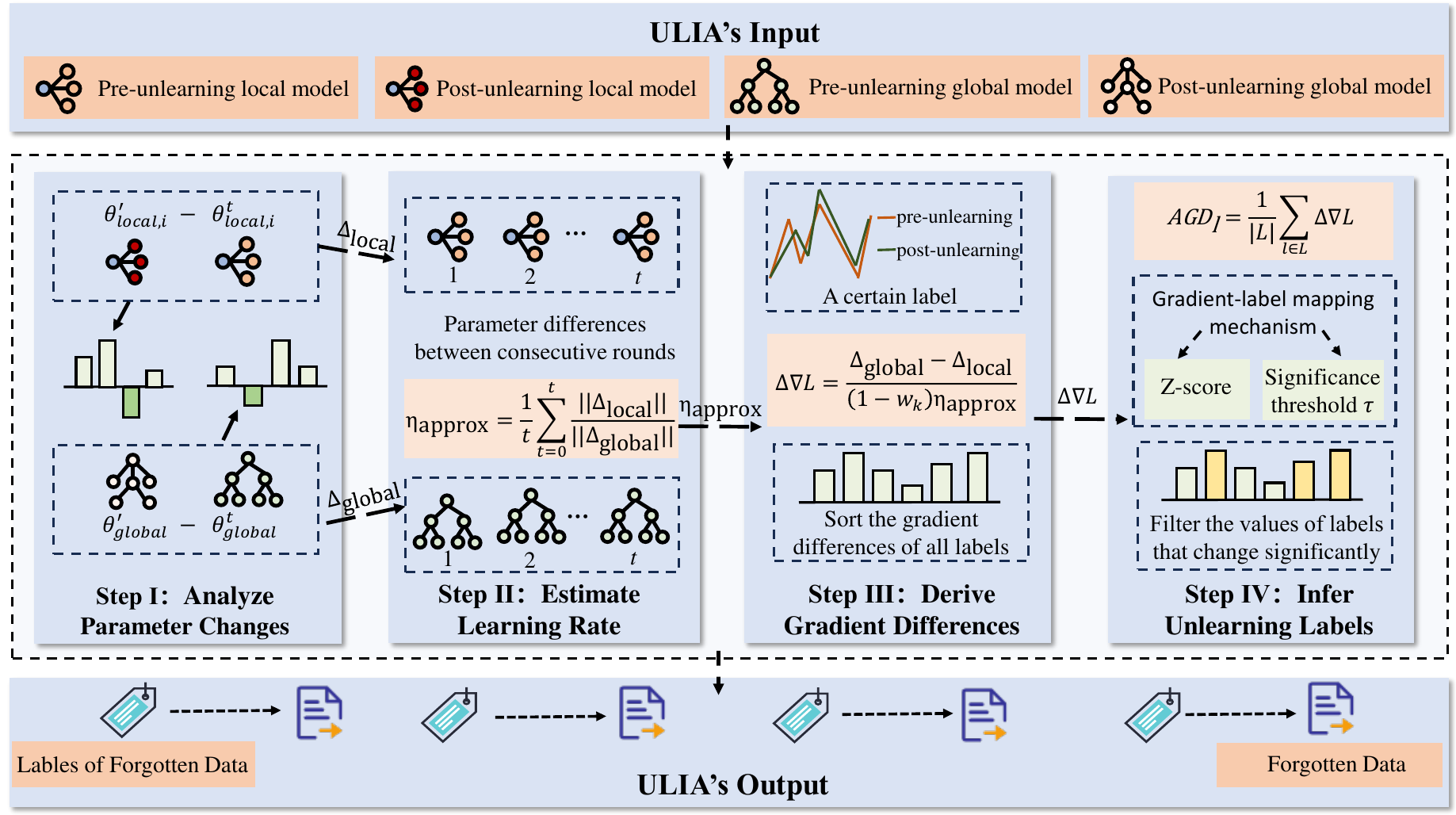}
\caption{Overview of Unlearning Label Inference Attack. It shows the four main steps: (1) analyzing parameter changes, (2) estimating the learning rate, (3) deriving gradient differences, and (4) inferring the forgotten labels. The diagram highlights how ULIA utilizes pre- and post-unlearning models to infer labels.}
\label{fig:Overview}
\end{figure}
\section{$\mathtt{ULIA}$: Label Inference Attack}
In this section, we propose $\mathtt{ULIA}$, a novel label inference attack that infer the labels of unlearning data across three FU levels: sample-level, class-level, and client-level, by analyzing the model differences in FU. Figure~\ref{fig:Overview} illustrates the proposed label inference attack in the context of FU. The implementation of the $\mathtt{ULIA}$ attack can be carried out in the following four steps.
\begin{itemize}
    \item \textit{Analysis of parameter changes.} This step compares local and global model parameters before and after unlearning to quantify the influence of forgotten data, providing a foundation for the attack. 
    \item \textit{Estimation of learning rate.} By analyzing parameter differences across multiple rounds, an effective learning rate is estimated to reconstruct gradients. 
    \item \textit{Derivation of gradient differences.} Using the estimated learning rate and parameter differences, approximate gradient differences are reconstructed. 
    \item \textit{Inferring the Labels of Forgotten Data.} By utilizing the gradient-label mapping mechanism, the degree of match between gradient changes and forgotten data labels is measured, enabling more accurate selection of labels.
\end{itemize}

\subsection{Analysis of Parameter Changes}
The first step in $\mathtt{ULIA}$ focuses on analyzing the changes in the model parameters caused by the unlearning process. By examining both the local and global model parameter changes, attackers can infer the influence of forgotten data on FL. Changes of local model parameters provide insights into how the removal of specific data affects individual clients, while changes of global model parameters reveal the aggregated impact of the unlearning operation across all clients.

Changes of local model parameters $\Delta_{\mathrm{local}}$ quantify the variation in the local model parameters of the target client before and after unlearning. This change, which captures the impact of removing the influence of forgotten data.
\begin{equation}
\Delta_{\mathrm{local}} = \theta_{\mathrm{local}, i}' - \theta_{\mathrm{local}, i}^{(t)}
\label{eq:4}
\end{equation}
where $\theta_{\mathrm{local}, i}^{(t)}$ represents the local model parameters before unlearning at the $t$-th communication round.
$\theta_{\mathrm{local}, i}'$ denotes the updated local model parameters.

Similarly, the global parameter change $\Delta_{\mathrm{global}}$, obtained from the model parameters before and after unlearning, is defined as:
\begin{equation}
\Delta_{\mathrm{global}} = \theta_{\mathrm{global}}' - \theta_{\mathrm{global}}^{(t)}
\label{eq:5}
\end{equation}
where $\theta_{\mathrm{global}}^{(t)}$ indicates the global model parameters before unlearning. $\theta_{\mathrm{global}}'$ refers to the updated global model parameters after unlearning.

The changes in the local model parameters $\Delta_{\mathrm{local}}$ reflect the adjustments made by the client during the update process to remove the influence of the unlearning data, revealing the contribution of specific data to the client's local model. By analyzing the changes in the global model parameters $\Delta_{\mathrm{global}}$ before and after unlearning, the extent of the influence of the unlearning data on the global model can be observed.


\subsection{Estimation of Effective Learning Rate}
The learning rate controls the step size of gradient updates in FL, directly affecting parameter changes. We estimate the learning rate by leveraging the differences between the local model parameters and the global model parameters from previous training rounds as given in Equation~(\ref{eq:6}). This enables $\mathtt{ULIA}$ to conduct attacks with more limited knowledge, enhancing its adaptability. 
\begin{equation}
\eta_{\mathrm{approx}} = \frac{1}{t} \sum_{t=0}^{t} \frac{\|\Delta_{\mathrm{local}}\|}{\|\Delta_{\mathrm{global}}\|}
\label{eq:6}
\end{equation}
where $\|\cdot\|_2$ denotes the $L2$-norm, $t$ is the number of rounds used for averaging.

The reasons for averaging over the previous $t$ rounds are as follows. Firstly, performing the averaging process helps mitigate the impact of noise and outliers in individual updates, leading to a more robust and reliable estimate. Secondly, it preserves the integrity of the gradient direction, as the learning rate influences only the magnitude of the update, not the direction. This  ensures that the estimated learning rate can be effectively used to derive gradient differences. By estimating an effective learning rate, attackers obtain the key parameters required to infer the gradients corresponding to the forgotten data.

\subsection{Derivation of Gradient Differences}

Attackers can infer the gradient updates associated with forgotten data by analyzing the difference between local and global parameter changes. This discrepancy provides insight into the contribution of the forgotten data to model optimization, forming a direct link between parameter variations and data characteristics. By leveraging an estimated learning rate, attackers can approximate the gradient difference introduced by the forgotten data.

In FL, the global model update at round \( t+1 \) follows the standard aggregation rule, as described in Equation~(\ref{eq:global_update}). When a client requests unlearning, its local model parameters are updated without the forgotten data, deviating from their original update path. The parameter changes before and after unlearning are defined as shown in Equations~(\ref{eq:4}) and~(\ref{eq:5}). Since the global model aggregates updates from all clients, its change can be rewritten as:
\begin{equation}
\Delta_{\text{global}} = \sum_{i=1}^{N} w_i \Delta_{\text{local}, i}.
\end{equation}
Computing the difference between global and local parameter changes gives:
\begin{equation}
\Delta_{\text{global}} - \Delta_{\text{local}} = (w_k - 1) \Delta_{\text{local}}.
\end{equation}
which reveals that the discrepancy between local and global updates is directly proportional to the forgotten data’s gradient contribution. Since the attackers do not directly observe \( \Delta \nabla L \), they approximate the discrepancy using the estimated learning rate \( \eta_{\text{approx}} \), yielding:
\begin{equation}
\Delta \nabla L = \frac{\Delta_{\text{global}} - \Delta_{\text{local}}}{(1 - w_k) \eta_{\text{approx}}}.
\end{equation}

This equation provides a mechanism for reconstructing the gradient updates influenced by the forgotten data. Given that gradient variations capture feature importance, the attackers can analyze the magnitude and sparsity of \(\Delta \nabla L\) to infer the forgotten data’s characteristics. By analyzing the approximated gradient differences, the attackers can identify which feature dimensions were most influenced by the forgotten data. Significant variations in specific gradient components indicate that these features were closely associated with the removed data. Furthermore, examining the concentration and distribution of gradient changes allows the attackers to assess the relative importance of individual features. 

The difference between local and global parameter updates provides a strong signal of the forgotten data’s impact on model training. As this discrepancy maintains a linear relationship with the forgotten data’s gradient contribution, it serves as a crucial indicator for reconstructing its characteristics. These findings reveal inherent weaknesses in existing FU mechanisms, demonstrating that even after explicit data removal, residual traces may still persist in the global model, posing potential privacy risks.

\subsection{Inferring the Labels of Forgotten Data}

The final step of the proposed method focuses on inferring the labels of forgotten data by analyzing the derived gradient differences. The unlearning operation introduces significant changes to the gradients associated with the forgotten data, making them more prominent in the analysis. By linking these gradient changes to the weights in the model's output layer, attackers can accurately infer the labels of the forgotten data. 
For each label category \( l \in L \), the average gradient difference is computed as:
\begin{equation}
\mathrm{AGD}_l = \frac{1}{|L|} \sum_{l \in L} \Delta\nabla L_l
\end{equation}
where \(\mathrm{AGD}_l\) represents the average gradient difference for a specific label category \( l \), and \(L\) is the set of all label categories.

This step emphasizes the innovative integration of gradient difference reconstruction with targeted label analysis, illustrating how the method effectively translates parameter and gradient variations into actionable insights, thereby enabling precise identification of the labels of forgotten data.

However, when multiple label categories are forgotten by the client, the server does not know the exact number of forgotten categories. Therefore, it cannot simply assume that the category with the largest gradient change accounts for all the forgotten categories. To address this, we propose a \textit{gradient-label mapping mechanism}, which establishes a one-to-one correspondence between gradient changes and the labels of forgotten data, and employs a dynamic filtering strategy to select labels that are more likely to correspond to the forgotten data. Specifically, this mechanism leverages two key methods: (1) It quantifies the changes induced by each label category during model parameter updates by utilizing the properties of the Z-score \cite{fei2021z}. 
\begin{equation}
Z_l = \frac{\mathrm{AGD}_l - \mu_{\mathrm{AGD}}}{\sigma_{\mathrm{AGD}}}
\end{equation}
where \(\mu_{\mathrm{AGD}}\) represents the average of gradient differences across all label categories, and \(\sigma_{\mathrm{AGD}}\) represents the standard deviation of the gradient differences across all label categories.
(2) It dynamically infers the number of forgotten data categories. After computing the Z-score for each label category, a predefined significance threshold $\tau$ is applied to filter a candidate set $L_{\text{candidate}}$ as given in Equation~(\ref{eq:13}), which includes the label categories most likely corresponding to the forgotten data of the client.

\begin{equation}
L_{\text{candidate}} = \{ l \mid Z_l > \tau\}
\label{eq:13}
\end{equation}

\section{Experiments}
\begin{table}[!t]
\small
\caption{Attack performance of $\mathtt{ULIA}$ applied to typical unlearning methods. The attack is performed under the condition that the quantity of forgotten samples accounts for $10\%$ of the client's total local data. $\mathcal{L}$ represents the number of categories of forgotten data labels.}
\label{tab1}
\centering
\renewcommand{\arraystretch}{0.9} 
\begin{tabular}{C{1.1cm}|C{0.5cm}|C{1.1cm}C{0.9cm}C{0.95cm}|C{1.1cm}C{0.9cm}C{0.95cm}|C{1.1cm}C{0.9cm}C{0.95cm}}
\cmidrule[1.5pt]{1-11}
\multicolumn{2}{c|}{\textbf{Methods}}& 
\multicolumn{3}{c|}{\textbf{FedEraser}} & 
\multicolumn{3}{c|}{\textbf{Rapid Retrain}} & 
\multicolumn{3}{c}{\textbf{SGA-EWC}} \\ 
\cmidrule(lr){3-5}
\cmidrule(lr){6-8}
\cmidrule(lr){9-11}

\multicolumn{1}{c}{\textbf{Datasets}} & 
\multicolumn{1}{c|}{\textbf{$\mathcal{L}$}} & 
\textbf{Sample} & \textbf{Class} & \textbf{Client} & \multicolumn{1}{c}{\textbf{Sample}} & 
\textbf{Class} & \textbf{Client} & \multicolumn{1}{c}{\textbf{Sample}} & 
\textbf{Class} & \textbf{Client} \\ 
\midrule

\multicolumn{11}{c}{\textbf{The attacker is aware of the number of label categories}} \\ 
\midrule

\multirow{3}{*}{\textbf{MNIST}} 
& 1 & 1.000 & 1.000 & 1.000 & 1.000 & 1.000 & 1.000  & 1.000 & 1.000 & 1.000 \\
& 2 & 0.970 & 1.000 & 1.000 & 0.970 & 1.000 & 1.000  & 0.940 & 1.000 & 1.000 \\
& 3 & 0.850 & 1.000 & 1.000 & 0.910 & 1.000 & 1.000  & 0.810 & 1.000 & 1.000 \\
\midrule
\multirow{3}{*}{\textbf{CIFAR}} 
& 1 & 0.970 & 1.000 & 1.000 & 1.000 & 1.000 & 1.000  & 0.950 & 1.000 & 1.000 \\
& 2 & 0.890 & 1.000 & 1.000 & 0.920 & 1.000 & 1.000  & 0.860 & 1.000 & 1.000 \\
& 3 & 0.800 & 1.000 & 1.000 & 0.860 & 1.000 & 1.000  & 0.780 & 1.000 & 1.000 \\
\midrule

\multicolumn{11}{c}{\textbf{The attacker does not know the number of label categories}} \\ 
\midrule

\multirow{3}{*}{\textbf{MNIST}} 
& 1 & 0.958 & 1.000 & 1.000 & 0.973 & 1.000 & 1.000 & 0.923 & 1.000 & 1.000  \\
& 2 & 0.898 & 1.000 & 1.000 & 0.925 & 1.000 & 1.000 & 0.855 & 1.000 & 1.000  \\
& 3 & 0.780 & 1.000 & 1.000 & 0.850 & 1.000 & 1.000 & 0.738 & 1.000 & 1.000  \\
\midrule
\multirow{3}{*}{\textbf{CIFAR}} 
& 1 & 0.919 & 1.000 & 1.000 & 0.924 & 1.000 & 1.000 & 0.887 & 1.000 & 1.000  \\
& 2 & 0.837 & 1.000 & 1.000 & 0.872 & 1.000 & 1.000 & 0.785 & 1.000 & 1.000  \\
& 3 & 0.747 & 1.000 & 1.000 & 0.817 & 1.000 & 1.000 & 0.692 & 1.000 & 1.000  \\

\midrule[1.5pt]
\end{tabular}
\end{table}

\subsection{Experimental Settings}
All experiments are performed on a workstation equipped with an Intel(R) Core(TM) i7-13700K processor, 64GB of RAM, and three NVIDIA RTX 4090 GPU cards.
$\mathtt{ULIA}$ is implemented using Python $3.8$ and PyTorch $2.1.0$.
\vspace{-0.3cm}
\subsubsection{Datasets and model.}
We select the following two datasets, which are widely used in FU. 1) \textbf{MNIST} \cite{lecun1998gradient}, a handwritten digit classification dataset consisting of $60,000$ training images and $10,000$ test images. 2) \textbf{CIFAR-10} \cite{krizhevsky2009learning}, a color image classification dataset containing $50,000$ training samples and $10,000$ test samples. We conduct experiments on both \textit{IID} and \textit{non-IID} data distributions, and perform training on the popular deep learning model \textbf{ResNet-18} \cite{he2016deep}.
\vspace{-0.3cm}
\subsubsection{FU methods.}
To comprehensively evaluate $\mathtt{ULIA}$, we apply it to the following three typical FU methods. 1) \textbf{FedEraser} \cite{liu2021federaser}, calibrates the historical updates of retained clients, enabling the server to efficiently reconstruct the global model. 2) \textbf{Rapid Retrain} \cite{liu2022right}, utilizes gradient and curvature information to identify more optimal descent directions, facilitating efficient retraining. 3) \textbf{SGA-EWC} \cite{wu2022federated}, proposes an efficient FU framework using reverse Stochastic Gradient Ascent (SGA) and Elastic Weight Consolidation (EWC) to quickly adjust model parameters and eliminate the influence of specific data.
\vspace{-0.3cm}
\subsubsection{Details of parameter settings.}
Before the unlearning process, FU is performed for $100$ rounds on $10$ clients, using a Stochastic Gradient Descent (SGD) optimizer with an initial learning rate of $0.01$, and a batch size of $64$. Additionally, we set the significance threshold as $\tau=2$ to more accurately infer the labels.
\vspace{-0.3cm}
\subsubsection{Evaluation metrics.}
Like previous works\cite{jiang2018acquisition,zheng2020distance}, we employ the widely-used Intersection over Union (IoU) method  to evaluate the ASR of $\mathtt{ULIA}$ in inferring the labels of the forgotten data. We perform 100 attack tests and calculate the average ASR of $\mathtt{ULIA}$. The ASR is calculated as follows:
\begin{equation}
\text{ASR} = \frac{|L_{\text{true}}^i \cap L_{\text{pred}}^i|}{|L_{\text{true}}^i \cup L_{\text{pred}}^i|}
\end{equation}
where \( |L_{true}^i \cap L_{pred}^i| \) represents the size of the intersection between the true label set \( L_{true}^i \) and the predicted label set \( L_{pred}^i \) for the \(i\)-th attack test. 

\begin{table*}[!t]
\caption{The impact of the quantity of forgotten samples on the ASR of $\mathtt{ULIA}$. $1\%$, $2\%$, and $5\%$ represent the percentage of samples requested to be forgotten, relative to the client's total local data.
}
\label{tab2}
\centering
\renewcommand{\arraystretch}{0.9} 
\small
\begin{tabular}{C{1.2cm}|C{0.6cm}|C{0.9cm}C{0.9cm}C{0.9cm}|C{0.9cm}C{0.9cm}C{0.9cm}|C{0.9cm}C{0.9cm}C{0.9cm}}
\cmidrule[1.5pt]{1-11}
\multicolumn{2}{c|}{\textbf{Methods}}& 
\multicolumn{3}{c|}{\textbf{FedEraser}} & 
\multicolumn{3}{c|}{\textbf{Rapid Retrain}} & 
\multicolumn{3}{c}{\textbf{SGA-EWC}} \\ 
\cmidrule(lr){3-5}
\cmidrule(lr){6-8}
\cmidrule(lr){9-11}

\multicolumn{1}{c}{\textbf{Datasets}} & 
\multicolumn{1}{c|}{\textbf{$\mathcal{L}$}} & 
\textbf{1\%} & \textbf{2\%} & \textbf{5\%} & \multicolumn{1}{c}{\textbf{1\%}} & 
\textbf{2\%} & \textbf{5\%} & \multicolumn{1}{c}{\textbf{1\%}} & 
\textbf{2\%} & \textbf{5\%} \\ 
\midrule

\multicolumn{11}{c}{\textbf{The attacker is aware of the number of label categories}} \\ 
\midrule

\multirow{3}{*}{\textbf{MNIST}} 
& 1 & 0.910 & 0.960 & 1.000 & 0.930 & 0.970 & 1.000 & 0.870 & 0.920 & 0.980 \\
& 2 & 0.820 & 0.850 & 0.900 & 0.850 & 0.890 & 0.920 & 0.760 & 0.810 & 0.880 \\
& 3 & 0.700 & 0.750 & 0.810 & 0.760 & 0.820 & 0.870 & 0.670 & 0.710 & 0.770 \\
\midrule
\multirow{3}{*}{\textbf{CIFAR}} 
& 1 & 0.830 & 0.890 & 0.940 & 0.850 & 0.920 & 0.980 & 0.770 & 0.850 & 0.900 \\
& 2 & 0.760 & 0.810 & 0.850 & 0.780 & 0.840 & 0.880 & 0.690 & 0.750 & 0.810 \\
& 3 & 0.680 & 0.720 & 0.770 & 0.700 & 0.760 & 0.810 & 0.650 & 0.690 & 0.740 \\
\midrule

\multicolumn{11}{c}{\textbf{The attacker does not know the number of label categories}} \\ 
\midrule

\multirow{3}{*}{\textbf{MNIST}} 
& 1 & 0.825 & 0.886 & 0.918 & 0.838 & 0.891 & 0.933 & 0.783 & 0.848 & 0.887 \\
& 2 & 0.783 & 0.833 & 0.872 & 0.827 & 0.868 & 0.907 & 0.697 & 0.745 & 0.803 \\
& 3 & 0.665 & 0.718 & 0.761 & 0.729 & 0.778 & 0.814 & 0.642 & 0.675 & 0.719 \\
\midrule
\multirow{3}{*}{\textbf{CIFAR}} 
& 1 & 0.778 & 0.847 & 0.902 & 0.793 & 0.853 & 0.916 & 0.747 & 0.806 & 0.843 \\
& 2 & 0.713 & 0.773 & 0.813 & 0.748 & 0.798 & 0.843 & 0.658 & 0.693 & 0.732 \\
& 3 & 0.648 & 0.691 & 0.735 & 0.662 & 0.714 & 0.759 & 0.623 & 0.648 & 0.668 \\

\midrule[1.5pt]
\end{tabular}
\end{table*}
\subsection{Attack Performance Evaluation}
We set the number of categories of forgotten data labels to $1$, $2$, and $3$, and perform attacks on three different FU methods at three levels, under two conditions: whether the attacker knows the number of forgotten data labels. Meanwhile, the quantity of forgotten samples is set to $10\%$ of the target client's data. The attack performance of $\mathtt{ULIA}$ applied to typical unlearning methods on the MNIST and CIFAR-10 datasets is reported in Table~\ref{tab1}.

$\mathtt{ULIA}$ achieves $100\%$ ASR for both class-level and client-level unlearning, because the large volume of forgotten data causes significant shifts in the model's parameters, making the inference easier. The changes in model weights and gradients are directly correlated with the forgotten data. By analyzing the gradient differences between the global model and local models, $\mathtt{ULIA}$ effectively identifies the forgotten data. For sample-level unlearning, the ASR gradually decreases as the number of forgotten data labels increases. However, on the MNIST dataset, when the number of forgotten label categories is $1$, the ASR reaches $95.8\%$ on FedEraser, $97.3\%$ on Rapid Retrain, and $92.3\%$ on SGA-EWC. This indicates that when the number of forgotten label categories is small, our attack still achieves strong performance, approaching $100\%$. Even with $3$ forgotten label categories, $\mathtt{ULIA}$ still achieves $73.8\%$ ASR. Similarly, on the CIFAR-10 dataset, $\mathtt{ULIA}$ maintains $69.2\%$ ASR under the same conditions. This difference is small compared to the ASR when attacker knows the number of forgotten data labels, indicating that $\mathtt{ULIA}$ is still able to effectively infer the labels even without complete information, demonstrating strong attack performance and adaptability.

\begin{table}[!t]
\small
\caption{Impact of the Non-IID Data Distribution. The attack is performed under the condition that the quantity of forgotten samples accounts for $10\%$ of the client's total local data.}
\label{tab3}
\centering
\renewcommand{\arraystretch}{0.9} 
\begin{tabular}{C{1.1cm}|C{0.5cm}|C{1.1cm}C{0.9cm}C{0.95cm}|C{1.1cm}C{0.9cm}C{0.95cm}|C{1.1cm}C{0.9cm}C{0.95cm}}
\cmidrule[1.5pt]{1-11}
\multicolumn{2}{c|}{\textbf{Methods}}& 
\multicolumn{3}{c|}{\textbf{FedEraser}} & 
\multicolumn{3}{c|}{\textbf{Rapid Retrain}} & 
\multicolumn{3}{c}{\textbf{SGA-EWC}} \\ 
\cmidrule(lr){3-5}
\cmidrule(lr){6-8}
\cmidrule(lr){9-11}

\multicolumn{1}{c}{\textbf{Datasets}} & 
\multicolumn{1}{c|}{\textbf{$\mathcal{L}$}} & 
\textbf{Sample} & \textbf{Class} & \textbf{Client} & \multicolumn{1}{c}{\textbf{Sample}} & 
\textbf{Class} & \textbf{Client} & \multicolumn{1}{c}{\textbf{Sample}} & 
\textbf{Class} & \textbf{Client} \\ 
\midrule


\multirow{3}{*}{\textbf{MNIST}} 
& 1 & 0.872 & 0.955 & 0.918 & 0.893 & 0.964 & 0.932 & 0.834 & 0.892 & 0.878  \\
& 2 & 0.828 & 0.914 & 0.858 & 0.852 & 0.934 & 0.872 & 0.775 & 0.848 & 0.812  \\
& 3 & 0.735 & 0.872 & 0.798 & 0.755 & 0.902 & 0.822 & 0.645 & 0.798 & 0.745  \\
\midrule
\multirow{3}{*}{\textbf{CIFAR}} 
& 1 & 0.812 & 0.914 & 0.868 & 0.835 & 0.925 & 0.882 & 0.765 & 0.864 & 0.825  \\
& 2 & 0.714 & 0.885 & 0.808 & 0.742 & 0.908 & 0.824 & 0.692 & 0.805 & 0.778  \\
& 3 & 0.625 & 0.834 & 0.752 & 0.652 & 0.845 & 0.782 & 0.585 & 0.748 & 0.695  \\

\midrule[1.5pt]
\end{tabular}
\end{table}
\vspace{-0.3cm}
\subsection{Sensitivity Evaluation}
In this subsection, we study the impact of three key factors, the quantity of forgotten samples, the non-IID data distribution and the significance threshold.

\vspace{-0.3cm}
\subsubsection{Impact of the Quantity of Forgotten Samples.}
The quantity of forgotten samples affects the magnitude of model parameter changes before and after unlearning. Therefore, we set the quantity of forgotten samples to $1\%$, $2\%$, and $5\%$ to evaluate $\mathtt{ULIA}$. The impact of the quantity of forgotten samples on the ASR of $\mathtt{ULIA}$ is shown in Table~\ref{tab2}.

The experimental results show that as the number of forgotten samples increases, the ASR of $\mathtt{ULIA}$ under different FU methods improves. Taking the MNIST dataset as an example, on FedEraser, when the number of forgotten label categories is $1$ and the forgotten sample percentage is $1\%$, the ASR of $\mathtt{ULIA}$ is $82.50\%$, while when the forgotten sample percentage is $5\%$, the ASR of $\mathtt{ULIA}$ is $91.83\%$. This is because as the number of forgotten samples increases, the gradient differences corresponding to the labels become more pronounced. However, even with $3$ forgotten label categories and only $1\%$ of the quantity of forgotten samples, $\mathtt{ULIA}$ can still achieve the ASR of $64.17\%$. Therefore, even in situations where the quantity of forgotten samples is small and the number of forgotten label categories is large, $\mathtt{ULIA}$ is still capable of effectively handling and adapting to these conditions.

\begin{figure}[htbp]
    \centering
    \begin{subfigure}[b]{0.8\textwidth}  
        \centering
        \includegraphics[width=\textwidth,height=0.4\textheight,keepaspectratio]{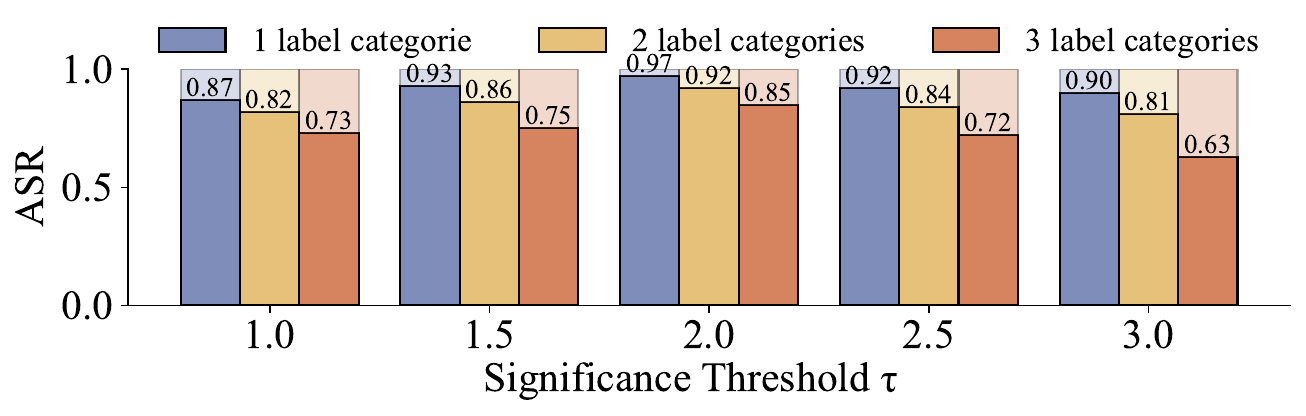}  
        \caption{On the MNIST dataset.}
        \label{fig:sub1}
    \end{subfigure}

    \begin{subfigure}[b]{0.8\textwidth}  
        \centering
        \includegraphics[width=\textwidth,height=0.4\textheight,keepaspectratio]{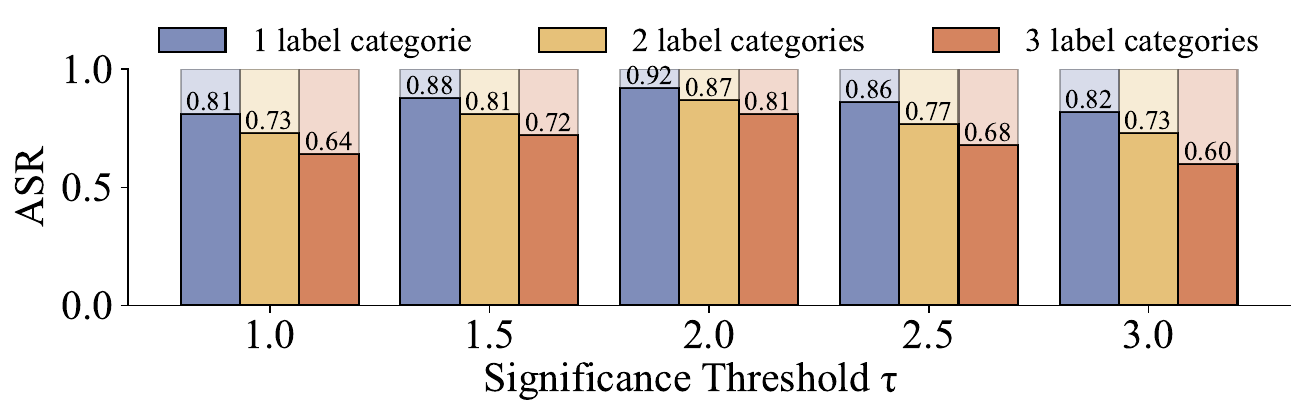}  
        \caption{On the CIFAR-10 dataset.}
        \label{fig:sub2}
    \end{subfigure}

    \vspace{-0.3cm} 
    \caption{The impact of different significance threshold on the ASR of $\mathtt{ULIA}$.}
    \vspace{-0.5cm} 
    \label{fig:main}
\end{figure}
\vspace{-0.3cm}
\subsubsection{Impact of the Non-IID Data Distribution.} The distribution of data across clients significantly influences the performance of $\mathtt{ULIA}$. In non-IID settings, the skewed data distribution causes uneven parameter updates, complicating the inference process. To evaluate the impact of non-IID distribution on $\mathtt{ULIA}$, we conduct experiments under varying levels of data heterogeneity. The effect on the ASR is shown in Table ~\ref{tab3}.

In the non-IID data environment, the ASR of $\mathtt{ULIA}$ is generally lower than that in the IID setting. This is due to the heterogeneity of the non-IID data distribution, which leads to large differences in data characteristics across clients, affecting the global model update process and making it more difficult to infer forgotten data. Nevertheless, $\mathtt{ULIA}$ still maintains a certain attack performance in the non-IID environment. On the MNIST dataset, the ASR ranges from 64.5\% to 96.4\%, while on the CIFAR-10 dataset, it ranges from 58.5\% to 92.5\%, demonstrating its strong adaptability and considerable attack performance.

\vspace{-0.3cm}
\subsubsection{Impact of Different Significance Threshold $\bm{\tau}$.}
The significance threshold $\tau$ is used to filter candidate forgotten label categories, and it has a significant impact on the ASR. Therefore, we performed experiments with different values of $\tau$, using $\mathtt{ULIA}$ applied in the Rapid Retrain method as an example. The experimental results on the MNIST dataset are shown in Figure~\ref{fig:main}(a), while the results on the CIFAR-10 dataset are shown in Figure~\ref{fig:main}(b). 

The experimental results indicate that as $\tau$ increases, the ASR first increases and then decreases. Around $\tau= 2$, $\mathtt{ULIA}$ achieves the best attack performance, regardless of whether the number of forgotten label categories is $1$, $2$, or $3$. Therefore, selecting an appropriate value of $\tau$ is crucial for the attack performance of $\mathtt{ULIA}$. If $\tau$ is set too high, some smaller but still important gradient changes may be missed, thus affecting the attack effectiveness. Conversely, if $\tau$ is set too low, too many label categories may be incorrectly inferred as forgotten categories, which could compromise the accuracy of the attack.

\vspace{-0.3cm}
\section{Conclusion}
In this paper, we have analyzed the label inference attacks against Federated Unlearning (FU). Our research has uncovered a significant privacy vulnerability within the FU framework. We have introduced $\mathtt{ULIA}$, a novel label inference attack that can infer the labels of unlearning data at three FU levels: sample-level, class-level, and client-level, by examining the model variations caused by FU. Our experiments show that $\mathtt{ULIA}$ demonstrates outstanding attack performance and adaptability. In future work, we will explore defense strategies on FU that protect against the label inference attacks, contributing to the development of more robust privacy-preserving techniques in FL systems.

\bibliographystyle{splncs04}  
\bibliography{references}     

\begin{thebibliography}{10}
\providecommand{\url}[1]{\texttt{#1}}
\providecommand{\urlprefix}{URL }
\providecommand{\doi}[1]{https://doi.org/#1}

\bibitem{arazzi2023blindsage}
Arazzi, M., Conti, M., Koffas, S., Krcek, M., Nocera, A., Picek, S., Xu, J.:
  Blindsage: Label inference attacks against node-level vertical federated
  graph neural networks. arXiv preprint arXiv:2308.02465  (2023)

\bibitem{che2023fast}
Che, T., Zhou, Y., Zhang, Z., Lyu, L., Liu, J., Yan, D., Dou, D., Huan, J.:
  Fast federated machine unlearning with nonlinear functional theory. In:
  International conference on machine learning. pp. 4241--4268. PMLR (2023)

\bibitem{fei2021z}
Fei, N., Gao, Y., Lu, Z., Xiang, T.: Z-score normalization, hubness, and
  few-shot learning. In: Proceedings of the IEEE/CVF International Conference
  on Computer Vision. pp. 142--151 (2021)

\bibitem{fu2022label}
Fu, C., Zhang, X., Ji, S., Chen, J., Wu, J., Guo, S., Zhou, J., Liu, A.X.,
  Wang, T.: Label inference attacks against vertical federated learning. In:
  31st USENIX Security Symposium (USENIX Security 22). pp. 1397--1414 (2022)

\bibitem{harding2019understanding}
Harding, E.L., Vanto, J.J., Clark, R., Hannah~Ji, L., Ainsworth, S.C.:
  Understanding the scope and impact of the california consumer privacy act of
  2018. Journal of Data Protection \& Privacy  \textbf{2}(3),  234--253 (2019)

\bibitem{he2016deep}
He, K., Zhang, X., Ren, S., Sun, J.: Deep residual learning for image
  recognition. In: Proceedings of the IEEE conference on computer vision and
  pattern recognition. pp. 770--778 (2016)

\bibitem{jiang2018acquisition}
Jiang, B., Luo, R., Mao, J., Xiao, T., Jiang, Y.: Acquisition of localization
  confidence for accurate object detection. In: Proceedings of the European
  conference on computer vision (ECCV). pp. 784--799 (2018)

\bibitem{jiang2020federated}
Jiang, J.C., Kantarci, B., Oktug, S., Soyata, T.: Federated learning in smart
  city sensing: Challenges and opportunities. Sensors  \textbf{20}(21), ~6230
  (2020)

\bibitem{jiang2024efficient}
Jiang, Y., Tong, X., Liu, Z., Ye, H., Tan, C.W., Lam, K.Y.: Efficient federated
  unlearning with adaptive differential privacy preservation. In: 2024 IEEE
  International Conference on Big Data (BigData). pp. 7822--7831. IEEE (2024)

\bibitem{krizhevsky2009learning}
Krizhevsky, A., Hinton, G., et~al.: Learning multiple layers of features from
  tiny images  (2009)

\bibitem{lecun1998gradient}
LeCun, Y., Bottou, L., Bengio, Y., Haffner, P.: Gradient-based learning applied
  to document recognition. Proceedings of the IEEE  \textbf{86}(11),
  2278--2324 (1998)

\bibitem{li2021label}
Li, O., Sun, J., Yang, X., Gao, W., Zhang, H., Xie, J., Smith, V., Wang, C.:
  Label leakage and protection in two-party split learning. arXiv preprint
  arXiv:2102.08504  (2021)

\bibitem{lin2024blockchain}
Lin, Y., Gao, Z., Du, H., Ren, J., Xie, Z., Niyato, D.: Blockchain-enabled
  trustworthy federated unlearning. arXiv preprint arXiv:2401.15917  (2024)

\bibitem{liu2021federaser}
Liu, G., Ma, X., Yang, Y., Wang, C., Liu, J.: Federaser: Enabling efficient
  client-level data removal from federated learning models. In: 2021 IEEE/ACM
  29th international symposium on quality of service (IWQOS). pp. 1--10. IEEE
  (2021)

\bibitem{liu2022clustering}
Liu, J., Lyu, X.: Clustering label inference attack against practical split
  learning. arXiv e-prints pp. arXiv--2203 (2022)

\bibitem{liu2022right}
Liu, Y., Xu, L., Yuan, X., Wang, C., Li, B.: The right to be forgotten in
  federated learning: An efficient realization with rapid retraining. In: IEEE
  INFOCOM 2022-IEEE Conference on Computer Communications. pp. 1749--1758. IEEE
  (2022)

\bibitem{liu2024survey}
Liu, Z., Jiang, Y., Shen, J., Peng, M., Lam, K.Y., Yuan, X., Liu, X.: A survey
  on federated unlearning: Challenges, methods, and future directions. ACM
  Computing Surveys  \textbf{57}(1),  1--38 (2024)

\bibitem{long2020federated}
Long, G., Tan, Y., Jiang, J., Zhang, C.: Federated learning for open banking.
  In: Federated learning: privacy and incentive, pp. 240--254. Springer (2020)

\bibitem{mothukuri2021survey}
Mothukuri, V., Parizi, R.M., Pouriyeh, S., Huang, Y., Dehghantanha, A.,
  Srivastava, G.: A survey on security and privacy of federated learning.
  Future Generation Computer Systems  \textbf{115},  619--640 (2021)

\bibitem{qiu2022your}
Qiu, P., Zhang, X., Ji, S., Du, T., Pu, Y., Zhou, J., Wang, T.: Your labels are
  selling you out: Relation leaks in vertical federated learning. IEEE
  Transactions on Dependable and Secure Computing  \textbf{20}(5),  3653--3668
  (2022)

\bibitem{shaik2024exploring}
Shaik, T., Tao, X., Xie, H., Li, L., Zhu, X., Li, Q.: Exploring the landscape
  of machine unlearning: A comprehensive survey and taxonomy. IEEE Transactions
  on Neural Networks and Learning Systems  (2024)

\bibitem{sheng2024robust}
Sheng, X., Bao, W., Ge, L.: Robust federated unlearning. In: Proceedings of the
  33rd ACM International Conference on Information and Knowledge Management.
  pp. 2034--2044 (2024)

\bibitem{su2023asynchronous}
Su, N., Li, B.: Asynchronous federated unlearning. In: IEEE INFOCOM 2023-IEEE
  Conference on Computer Communications. pp. 1--10. IEEE (2023)

\bibitem{voigt2017eu}
Voigt, P., Von~dem Bussche, A.: The eu general data protection regulation
  (gdpr). A Practical Guide, 1st Ed., Cham: Springer International Publishing
  \textbf{10}(3152676),  10--5555 (2017)

\bibitem{wang2022federated}
Wang, J., Guo, S., Xie, X., Qi, H.: Federated unlearning via
  class-discriminative pruning. In: Proceedings of the ACM Web Conference 2022.
  pp. 622--632 (2022)

\bibitem{wang2024efficient}
Wang, Z., Gao, X., Wang, C., Cheng, P., Chen, J.: Efficient vertical federated
  unlearning via fast retraining. ACM Transactions on Internet Technology
  \textbf{24}(2),  1--22 (2024)

\bibitem{wu2022federated}
Wu, L., Guo, S., Wang, J., Hong, Z., Zhang, J., Ding, Y.: Federated unlearning:
  Guarantee the right of clients to forget. IEEE Network  \textbf{36}(5),
  129--135 (2022)

\bibitem{yuan2023federated}
Yuan, W., Yin, H., Wu, F., Zhang, S., He, T., Wang, H.: Federated unlearning
  for on-device recommendation. In: Proceedings of the sixteenth ACM
  international conference on web search and data mining. pp. 393--401 (2023)

\bibitem{zhang2023fedrecovery}
Zhang, L., Zhu, T., Zhang, H., Xiong, P., Zhou, W.: Fedrecovery: Differentially
  private machine unlearning for federated learning frameworks. IEEE
  Transactions on Information Forensics and Security  (2023)

\bibitem{zhang2022data}
Zhang, X., Zhou, X., Chen, K.: Data leakage with label reconstruction in
  distributed learning environments. In: International Conference on Machine
  Learning for Cyber Security. pp. 185--197. Springer (2022)

\bibitem{zhao2020idlg}
Zhao, B., Mopuri, K.R., Bilen, H.: idlg: Improved deep leakage from gradients.
  arXiv preprint arXiv:2001.02610  (2020)

\bibitem{zhao2023federated}
Zhao, Y., Wang, P., Qi, H., Huang, J., Wei, Z., Zhang, Q.: Federated unlearning
  with momentum degradation. IEEE Internet of Things Journal  (2023)

\bibitem{zheng2020distance}
Zheng, Z., Wang, P., Liu, W., Li, J., Ye, R., Ren, D.: Distance-iou loss:
  Faster and better learning for bounding box regression. In: Proceedings of
  the AAAI conference on artificial intelligence. vol.~34, pp. 12993--13000
  (2020)

\end{thebibliography}

\end{document}